\documentclass[5p]{elsarticle}
\usepackage{amsmath,amssymb,amsfonts}
\usepackage[utf8]{inputenc}
\usepackage{graphicx}
\usepackage{todonotes}
\usepackage[bookmarks=false]{hyperref}
\usepackage[author={Max Schlepzig}]{pdfcomment}
\hypersetup{colorlinks=false,pdfborder={0 0 0},hypertexnames=false}
\usepackage{algorithm}
\usepackage[noend]{algpseudocode}
\usepackage{subfig}
\usepackage{textgreek}
\usepackage{gensymb}
\usepackage[thinc]{esdiff}

\newif\ifRoadmap
\Roadmaptrue 
\Roadmapfalse

\newcommand{\orcid}[1]{\href{https://orcid.org/#1}{\includegraphics*[width=10pt]{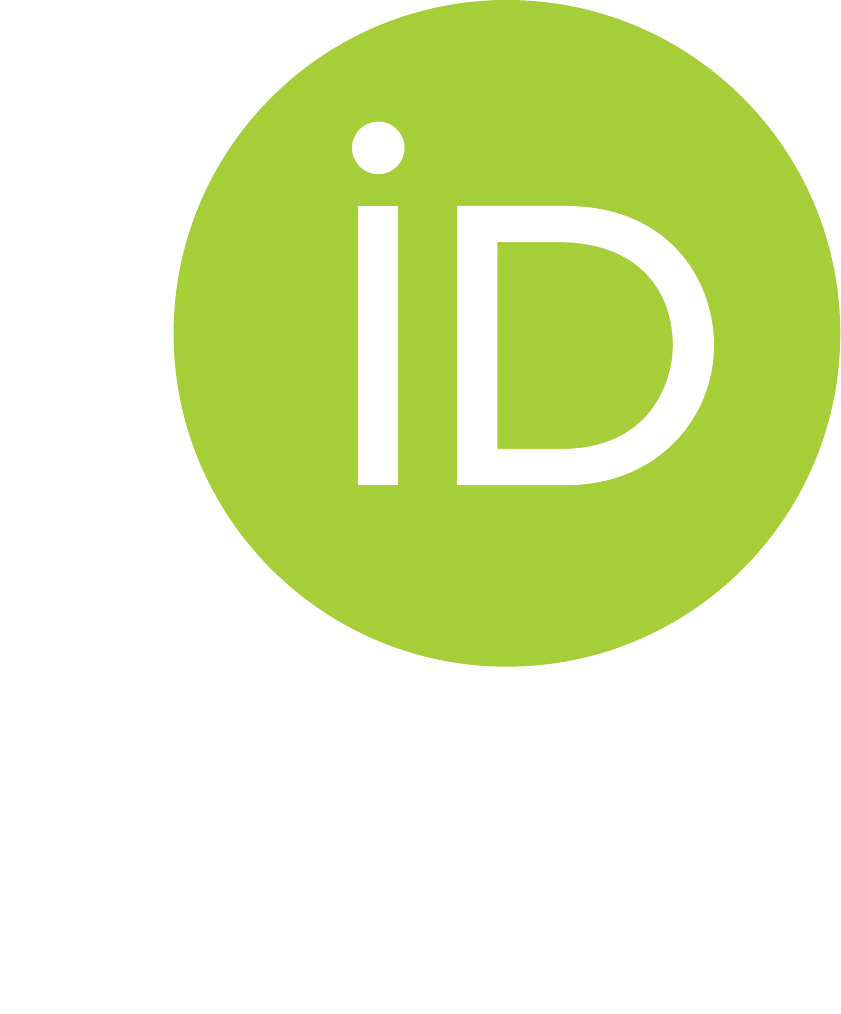}}}

\begin{document}
%
\title{Event-based sampled ECG \\ morphology reconstruction through self-similarity}
%
%
%

\author[1]{Silvio~Zanoli\orcid{0000-0002-0316-1657}}
\author[1]{Giovanni~Ansaloni\orcid{0000-0002-8940-3775}}
\author[2]{Tomás~Teijeiro\orcid{0000-0002-2175-7382}}
\author[1]{David~Atienza\orcid{0000-0001-9536-4947}}

\address[1]{Embedded Systems Laboratory (ESL), École Polytechnique Fédérale de Lausanne (EPFL), 1015 Lausanne}
\address[2]{Department of Mathematics, University of the Basque Country (UPV/EHU), Bilbao, Spain}


%
%

\markboth{...}%
{Zanoli \MakeLowercase{\textit{et al.}}: Event-based sampled ECG \\ morphology reconstruction through self-similarity}
%




\begin{abstract}
\textit{Background and Objective:} Event-based analog-to-digital converters allow for sparse bio-signal acquisition, enabling local sub-Nyquist sampling frequency. However, aggressive event selection can cause the loss of important bio-markers, not recoverable with standard interpolation techniques. 
In this work, we leverage the self-similarity of the electrocardiogram (ECG) signal to recover missing features in event-based sampled ECG signals, dynamically selecting patient-representative templates together with a novel dynamic time warping algorithm to infer the morphology of event-based sampled heartbeats. \\
\textit{Methods:} We acquire a set of uniformly sampled heartbeats and use a graph-based clustering algorithm to define representative templates for the patient. Then, for each event-based sampled heartbeat, we select the morphologically nearest template, and we then reconstruct the heartbeat with piece-wise linear deformations of the selected template, according to a novel dynamic time warping algorithm that matches events to template segments.\\
\textit{Results:} Synthetic tests on a standard normal sinus rhythm dataset, composed of approximately 1.8 million normal heartbeats, show a big leap in performance with respect to standard resampling techniques. In particular (when compared to classic linear resampling), we show an improvement in P-wave detection of up to 10 times, an improvement in T-wave detection of up to three times, and a 30\% improvement in the dynamic time warping morphological distance.\\
\textit{Conclusion:} In this work, we have developed an event-based processing pipeline that leverages signal self-similarity to reconstruct event-based sampled ECG signals. Synthetic tests show clear advantages over classical resampling techniques.
\end{abstract}
\maketitle

%

\section{Introduction}
\label{sec:introduction}

Nowadays, heightened life expectancy and unhealthy lifestyles make chronic diseases, and in particular chronic heart diseases, the leading cause of death worldwide~\cite{cardio_diseases_leading_death_cause}.
Such conditions are long-lasting and not extensively observable inside hospitals, both for the short time of observation and the restricted set of activities a patient can do while hospitalized. Moreover, the need for a long observation period requires non-invasive solutions that impact the patient life as little as possible. This requirements make wearable solutions key for chronic disease monitoring. 

One of the primary concerns of wearables is energy efficiency, as their main requirement is to function for the longest possible time while being unobtrusive to the patient. The energy budget of any battery operated device can be divided into four main categories: computation, storage, communication, and data acquisition. While computation, communications, and storage have been greatly studied and optimized in recent years~\cite{Wei19, Abadal20, Panades20, Pullini18}, data acquisition remains a field scarcely explored. Nonetheless, the energy budget in modern wearable systems is highly affected by the signal acquisition component~\cite{rincon2011development}.

In their seminal work on sampling, Nyquist and Shannon defined an upper bound to the sampling rate, called Nyquist frequency~\cite{Shannon1949}, which is two times the maximum spectral component of the signal under analysis. Signals acquired following the Nyquist-Shannon theorem are called uniformly sampled. However, it is possible to define many sampling schemes that do not rely on a uniform spacing between samples. 
One of the main non-uniform sub-Nyquist sampling approach is event-based (EB) sampling: a data-acquisition strategy that aims to record signal points only when certain events happen in the signal.

In recent years, two main event-based sampling techniques have emerged: 1) level-crossing analysis \cite{level_crossing}, and 2) polygonal approximation~\cite{polygonal_approximator}, which have shown to be able to greatly reduce the average sampling frequency of a signal up to 90\%~\cite{polygonal_approximator}. However, as the reduction rate increases, the signal fidelity greatly degrades, since all these acquisition approaches are lossy by their very nature.

We depart from the aforementioned sampling methods, considering the acquired events as fiducial points of the recorded signal. We then use these points, together with a set of representative signal templates, to drive a novel approach to signal reconstruction. 

Our approach is applicable to signals that are representable by a set of templates, such as the electrocardiogram (ECG)~\cite{context_based_QRS_clustering}. Indeed, since the ECG is the recording of the electrical polarization of the muscular tissue of the heart~\cite{SORNMO2005411}, only a finite amount of ECG morphologies can physically exist~\cite{heartbeat_morphologies}. These morphologies are characterized by the presence, absence, and shape of three main complexes, representing three polarization and depolarization phases of the cardiac muscle: P-wave, QRS complex, and T-wave. Hence, we can consider an ECG recording as the non-exact repetition of a set of patient-specific heartbeats (i.e.: templates), where every repetition has a degree of deformation with respect to a selected template.

\begin{figure*}[!t]
	\centering
    \includegraphics[width=.7\textwidth]{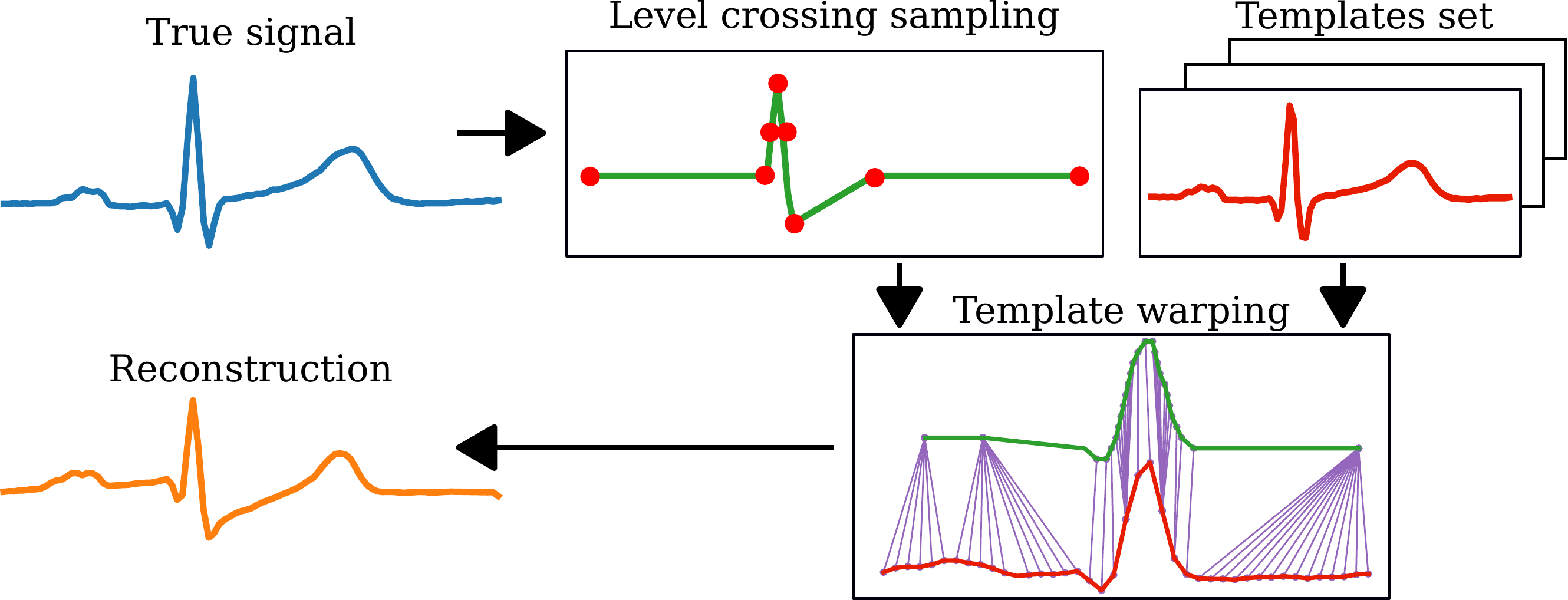}
	\caption{ECG signal reconstruction processing pipeline: each heartbeat is first sampled with a level crossing ADC, then compared to a set of templates. The most similar template is then warped to better represent the recorded events.}
	\label{fig:sys_overview}
\end{figure*}

We embodied this approach in the processing pipeline shown in Fig.~\ref{fig:sys_overview},  where we reconstruct an ECG signal sampled using level crossing~\cite{level_crossing}. To achieve this, we leverage the ECG self similarity~\cite{fractal_ECG_reconstruction} and a novel customization of the dynamic time warping algorithm (DTW)~\cite{DTW}, that we called Information-Injected Differential Dynamic Time Warping (II-DDTW). First, the presented system uniformly acquires a set of heartbeats (templates) representative of the signal. Then, for every EB sampled heartbeat, it uses II-DDTW to both find the best fitting template and optimally deform it so as to pass through all the recorded events. Moreover, whenever the templates set is not anymore representative of the underlying signal, the processing pipeline acquires a new templates set.

We test the developed processing pipeline against the reference MIT-BIH normal sinus rhythm database~\cite{MIT-BIH}.
This database constrains the variability in the input by considering only normal rhythm situations, allowing us to compare our method with standard resampling techniques, while focusing on the key aspects of signal reconstruction.

The reconstructed signals obtained with our approach are evaluated using three merit figures: 1) an element-wise aggregated distance (Percentage root mean square difference), which measures overall signal correctness, 2) a morphological distance (DTW distance), measuring the similarities between the compared heartbeats, and 3) P/T wave delineation accuracy, an application specific measures that shows if ECG relevant structures are present and rightly positioned. These three measures, spanning from signal agnostic to signal specific, give a comprehensive view  of the reconstruction accuracy. The results are then compared to the results obtained by reconstructing the signal with three different methods: sample\&hold, linear resampling, and SP-line resampling.

Hence, the main outcomes of our work are:
\begin{itemize}
    \item The relevance of self-similarity in EB-sampled signals.
    \item The possibility, for EB-sampled signals, to be represented by a templates set.
    \item The morphological mapping between templates and EB-signals, through a novel strategy called II-DDTW.
\end{itemize}

\section{Background work}
\label{sec:background}

In this section, we first explore the core concepts we leverage in our work. Then, we review the works most similar to ours, highlighting the key differences and the concepts we ourselves took as inspiration.


\subsection{Foundation notions}
\label{subsec:fundamental}

Our methodology relies on three core concepts: 1) self-similarity, 2) event-based sampling, and 3) dynamic time warping (DTW).

\textbf{Self-similarity:}
We define a signal to be self-similar when it can be approximately described by a function of a subsection of the signal itself. The term "self-similar" appears in other disciplines and works as a characteristic of a fractal-type mathematical object \cite{Mandelbrot_1985}. The self similarity aspect of the ECG signal is used in~\cite{fractal_ECG_reconstruction}. In their work, \cite{fractal_ECG_reconstruction} design a compression system for ECG signals based on multi-scale analysis, where each heartbeat is encoded as the scaling and rotations of prototype heartbeats.

\textbf{Event-based sampling:}
Event-based sampling is the action of drawing the value of a measured signal only when it has a predetermined behaviour (an event), using an event-based analog to digital converter (EB-ADC).

To represent a signal using only its samples requires a set of conditions on the sampling function. Most famously, the Nyquist-Shannon theorem~\cite{Shannon1949} imposes a uniform-grid sampling frequency of two times the maximum spectral content of the signal under analysis. Other approaches~\cite{non_uniform_sampling_reconstruction_conditions,sub_nyquist_optimal} broaden this condition while still achieving a loss-less reconstruction of the sampled function. However, these approaches do not use any contextual information to determine whether a signal section conveys important features or not. Instead, the approach here analyzed defines an event as a logical condition that can be verified based on the signal behavior, and such events are then used to trigger the sampling process. 

The two most prominent examples of EB-sampling are level-crossing analysis
and polygonal approximation.
In level-crossing analysis~\cite{level_crossing} an event is generated every time the measured physical quantity cross a set of levels. Such operation can either be analog~\cite{analog_level_crossing}, where a sampler detects the crossing of analog-defined levels or digital~\cite{level_crossing}, where the signal is digitally acquired by a standard analog-to-digital converter (ADC) with the addition of a custom logic that defines a set of digital levels and forwards the samples to the main processing unit only when such levels are crossed. Instead, in polygonal approximation~\cite{polygonal_approximator} the signal is digitally acquired by a standard ADC but with the addition of a custom logic that forwards the samples to the main processing unit only when the error between the uniformly sampled signal and its linear approximation grows bigger than a threshold $\varepsilon$.

Moreover, EB sampling is related to technologies such as Compressed Sensing~(CS)~\cite{compressed_sensing}, which achieve data compression through two signal-agnostic measurement matrices, making the acquired signal sparse. However, contrary to EB-sampling, in CS the length of the sparse representation is signal independent.


\textbf{Dynamic time warping:}
DTW~\cite{DTW} is an algorithm that takes as input two vectors, not necessarily with the same dimensionality, and  outputs a distance between the two, together with a vector matching sequence. Such sequence, also called edit path, associates every point in the first vector with one or more points of the second vector. This algorithm is of particular interest to this work since it has proven to be effective in the alignment of ECG recordings~\cite{DTW_on_ECG} and because of the interpretability of the output distance as a morphological difference between the two input vectors. Formally, given two vectors $v_1$ and $v_2$ with dimensions $N$ and $M$, the DTW algorithm computes a matrix $D$ following Eq.~\ref{eq:DTW}:
\begin{equation}
\label{eq:DTW}
    D_{i,j} = |v_1[i] - v_2[j]| + min(D_{i,j-1},D_{i-1,j-1}D_{i-1,j})
\end{equation}

The boundary conditions on the $D$ matrix computation vary depending on the priors the user might define. However, they are typically defined as $D_{0,0} = 0,~D_{0,j} = D_{i,0} = \infty$~\cite{DTW}. The warping path is computed starting from $P_0=(N,M)$ and selecting the next point in the path according to Eq.~\ref{eq:DTW_path}:
\begin{equation}
\label{eq:DTW_path}
\begin{split}
     P_{l-1} &= (i,j)\\
     P_l &= argmin_{i,j,i-1,j-1} \{D_{i-1,j-1},D_{i,j-1},D_{i-1,j} \}
\end{split}
\end{equation}

The process is iteratively repeated until $D_{0,0}$ is reached. The list $P_0,...,P_l$ is then a sequence of tuples that associates each point in $v_1$ to one or more points in $v_2$ and vice versa, keeping the order relation between samples and warping the two signals by elongating sections of the two vectors. Finally, $D_{N,M}$ is called the DTW distance and can be interpreted as a measurement of morphological distance, as it increases with the path length and with the distances in each matched point.
DTW computes the morphological distance between two vectors, and the warping path that minimizes this distance. 
In this work, we use these properties to compute the pertinence of a template to an EB-sampled beat, and then to match each event.

The general formulation of the DTW algorithm shows some practical issues when dealing with signals without prominent matching features or with missing signal samples (as it is the case for event-based signals). The most relevant problem is the singularity problem~\cite{DTW,ddtw} where multiple points of one vector get (wrongly) associated with one single point of the other.
Moreover, the singularity problem exacerbates, in extreme cases, into the information mapping problem, where two non-matching features (where a feature is a distinct set of points not caused by noise) between $v_1$ and $v_2$ get mapped together. 

Several techniques have been developed to assess this problem like distance matrix step pattern re-definition~\cite{step_patterns_DTW} and differential dynamic time warping~\cite{ddtw}. In particular, this last technique consists in changing the distance function presented in Eq. \ref{eq:DTW} to the one in Eq. \ref{eq:DDTW} to include the difference in derivative instead of in value, thus capturing better a morphology change instead of the specific value.

\begin{equation}
\label{eq:DDTW}
\begin{split}
    D_{i,j} = |\frac{v_1[i]-v1[i-1]}{t1[i]-t1[i-1]} - \frac{v_2[j]-v2[j-1]}{t2[j]-t2[j-1]}| + \\ min(D_{i,j-1},D_{i-1,j-1}D_{i-1,j})
\end{split}
\end{equation} 

Finally, a modern approach we took as an example and point of departure for our work is the technique developed in~\cite{event_DTW}, named Event-Based Dynamic Time Warping (EB\-DTW). This approach starts by pre-processing the vector, identifying ascending and descending slopes sections\footnote{These slopes are called events in~\cite{event_DTW}.} in the two vectors. Then, it matches the events between the two vectors before the DTW algorithm. The distance in Eq.~\ref{eq:DTW} is then complemented by the insertion of the constraint that distances are computed only between points that are in the matching slopes.

This last technique is of interest to our application as it introduces the idea of injecting prior information in the computation of the DTW algorithm to improve its performance. However, such a strategy is not a viable solution for our problem, as entire slopes can be removed by the event-based sampling in EB-ECG signals. Moreover, while the technique in~\cite{event_DTW} is used for matching the same signal recorded by two different methods, in our work we couple a template with any EB-sampled heartbeat. Hence, we can not assume having the same features (in this case slopes) between the two compared signals. To overcome these issues, we introduce in Section~\ref{sec:methods} the concept of Information-Injected Differential Dynamic Time Warping (II-DDTW).


\subsection{Literature review}
\label{subsec:review}


The objective of our work is ECG signal reconstruction from a reduced set of key samples. This can be interpreted as the inverse problem to the signal compression task. However, classic decompression techniques \cite{compression_algorithms} are strictly coupled with the relative compression method. Our work, while requiring a specific class of signal, does not need a specific EB-sampling method.

Another related line of study can be identified in ECG representational studies. In~\cite{Self_organize_maps_clustering,context_based_QRS_clustering,representative_morphologies} the authors represent full ECG recordings through their most representative heartbeats, clustering each beat so to present a comprehensive view of a patient through key examples. While this approach shares similar intuitions with our work, we use the representative heartbeats for a different purpose, that of  templates for heartbeat reconstruction. 

Moreover, contrary to the aforementioned works, we determine the representative heartbeats through a graph clustering algorithm~\cite{Graph_clustering} called affinity propagation~\cite{Affinity_propagation} to solve both the uncertainty in the number of clusters and the non-convexity of the formulated clustering problem.

Finally, EB-sampling has seen a significant rise in interest in recent years \cite{level_crossing,analog_level_crossing,polygonal_approximator}. However, signal reconstruction has always been considered a secondary task, assessed using standard interpolation techniques like linear or SP-line interpolation~\cite{interpolations}. Here, we focus on optimizing signal reconstruction in those instances of EB-sampling leading to high sampling reduction factors.

\begin{figure}[!t]
	\centering
    \includegraphics[width=0.95\linewidth]{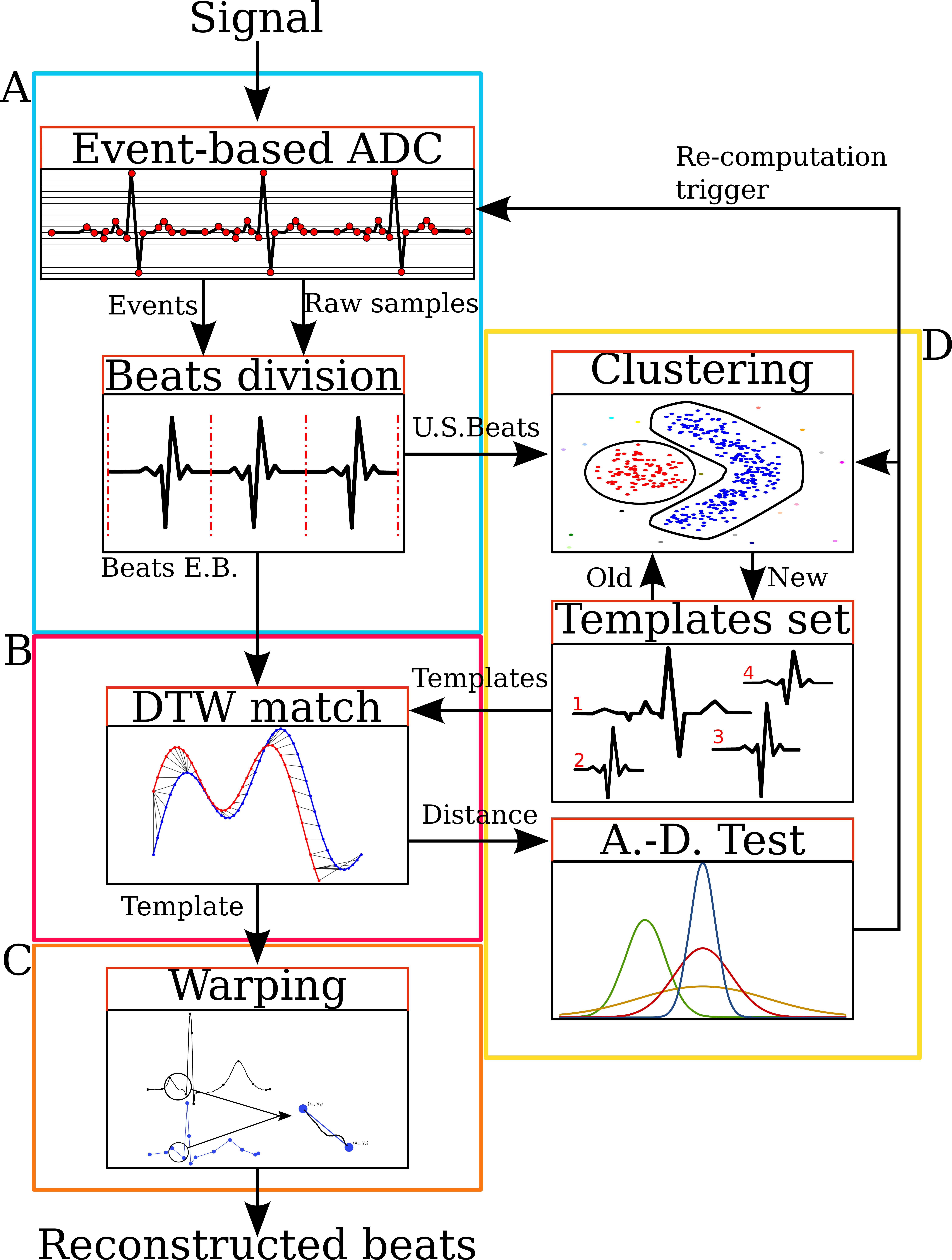}
    
	\caption{A bird-eye view of the system described in Section~\ref{sec:methods}. Each labeled box corresponds to specific subsections. A: Acquisition (Section~\ref{subsec:acquisition}), B: Differential dynamic time warping with information injection (Section~\ref{subsec:DTW}), C: Template-based reconstruction (Section~\ref{subsec:reconstruction}), D: Templates computation (Section~\ref{subsec:templates_computation}).}
	\label{fig:system}
\end{figure}



\section{Methodology}
\label{sec:methods}
A high-level view of the proposed methodology is shown in Fig.~\ref{fig:system}, divided into functional blocks.
The reconstruction methodology here presented works on a beat-by-beat basis. Our framework algorithmically selects representative heartbeats as templates. It then uses them to reconstruct event-based ECG acquisitions, warping templates using the acquired events as fiducial points, i.e. forcing the warped template to pass through them.  

Section \ref{subsec:acquisition} details the block in Fig.~\ref{fig:system}-A, where the system EB-samples the ECG signal, mark the QRS complexes, and uses this information to extract EB-heartbeats. 

Then, Section \ref{subsec:DTW} describes Fig.~\ref{fig:system}-B, where the II-DDTW algorithm (a pattern-matching algorithm) compares each new EB-heartbeat with a set of templates, and selects an optimal template.

Section \ref{subsec:reconstruction} delineates the block in Fig.~\ref{fig:system}-C, where EB-heartbeats are reconstructed based on the warping path defined by the previous block.

Finally, Section \ref{subsec:templates_computation} characterize Fig.~\ref{fig:system}-D, where the templates set is updated in order to keep the set relevant to the current EB-sample signal.

\subsection{Acquisition}
\label{subsec:acquisition}


The starting point for reconstruction is a sequence of samples acquired by a LC-ADC. Samples are represented by the sequence of time-value tuples $(t,v)$: the time of acquisition and the threshold value being crossed.

In Fig.~\ref{fig:system}-A, the system EB-samples the ECG signal using a level-crossing ADC such as the one described in~\cite{level_crossing}. Then, it marks the QRS complexes and extracts EB-heart-beats. This can be done, as previously shown in \cite{EB_QRS}, by re-implementing the  gQRS-detection algorithm \cite{physioToolkit} to work with EB-sampled signals.
Using the QRS timing information, the processing pipeline computes the instantaneous RR interval and uses it to define the heartbeats boundaries.
Finally, this processing pipeline section also provides the uniformly sampled heartbeats to the templates set recomputation block, whenever the current set is not anymore representative of the EB-heartbeats.

\subsection{Differential dynamic time warping with information injection}
\label{subsec:DTW}

We here introduce a novel algorithm, named Information Injected Differential Dynamic Time Warping (II-DDTW), which is an evolution of the DTW algorithm described in Section \ref{subsec:fundamental}. We use this novel approach to find the best fitting template for a EB-sampled heartbeat, and compute the warping parameters for the chosen template.

\textbf{Information-Injected DDTW.}
To address the DTW problems mentioned in Section~\ref{subsec:fundamental}, first, we opt to use the differential approach to the DTW algorithm (DDTW), expressed in Eq.~\ref{eq:DDTW}. Then, we re-formulated the DDTW accumulated distance to take into account additional information and use it to guide the warping process. The developed formulation for Information-Injected DDTW (II-DDTW) is shown in Eq.~\ref{eq:II-DDTW}, where the $t_1$, and $t_2$ terms represent the aforementioned additional information.
\begin{equation}
\label{eq:II-DDTW}
\begin{split}
    D_{i,j} = (1+\lambda|t_1[i]-t_2[i]|)\cdot|\diff{v_1[i]}{t_1} - \diff{v_2[j]}{t_2}| + \\ min(D_{i,j-1},D_{i-1,j-1}D_{i-1,j})
\end{split}
\end{equation}

The motivation of such a formulation can be found in the very nature of our study. EB-sampled signals are composed of data points represented by a tuple: (time, value).
We can define a time base also for uniformly sampled signals using the sampling period.
The time difference information leads the mapping between events and templates, enforcing a loose time matching requirement.


Moreover, since we can not ensure template and EB-sampled heartbeats to have the same duration, we normalize the timescale for both vectors to 1. We use the differential version of the DTW distance to better capture the changes in morphologies. Finally, we introduce a correction factor $\lambda$ to modify the behavior of the algorithm, giving more weight to the accumulated distance, as per Eq.~\ref{eq:DDTW}, or the additional (injected) information, in our case time, resulting in the multiplication factor in Eq.~\ref{eq:II-DDTW}.

\subsection{Template based reconstruction}
\label{subsec:reconstruction}
Using the definition of Section~\ref{subsec:review}, sections of the ECG signal are represented by morphological deformations of templates. Given an EB-heartbeat, the II-DDTW block (Fig.~\ref{fig:system}-B) computes the II-DDTW between the EB signal and all the elements of the templates set. Then, it selects the template with the smallest distance as representative.

\begin{figure}[!t]
	\centering
    \includegraphics[width=1\linewidth]{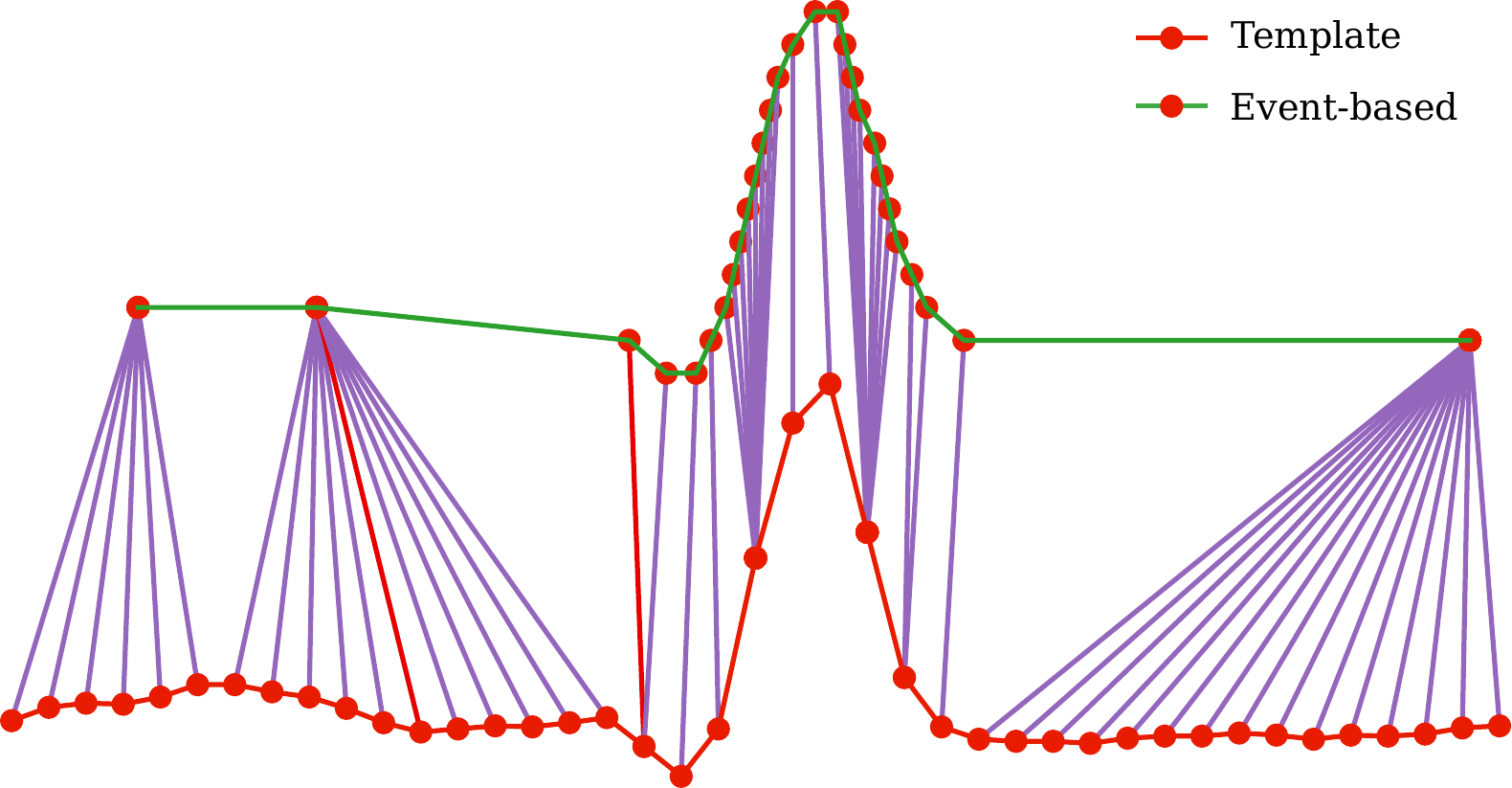}
    
	\caption{Example of an EB sampled heartbeat coupled with a template. The warping path is represented here by the purple lines, connecting events with the corresponding points in the template. The red lines exemplify the middle point in the points set associated with the corresponding event.}
	\label{fig:template-matching}
\end{figure}

To warp the selected representative template, the warping block (Fig.~\ref{fig:system}-C) uses the warping path to match every couple of consecutive events to a set of corresponding points in the template. As shown in Fig.~\ref{fig:template-matching}, to define the template segment corresponding to the each EB tuple the warping block uses the central half of the points assigned to the events: half of the points associated with the first event, and half of the points associated with the second event. This process produces a set of uniformly sampled segments (coming from the selected template), each associated with a specific segment in the EB sampled heartbeat.   

The reconstruction is then performed by deforming each template segment such that the edges of the segment match the two corresponding EB samples. First, the EB segments are shifted such that the first element is $(0,0)$, as described in Eq.~\ref{eq:shift_segment}:

\begin{equation}
\begin{split}
\label{eq:shift_segment}
    E_{i_s}[j] = E_i[j]-E_i[0] \\
    T_{i_s}[j] = T_i[j]-T_i[0]
\end{split}
\end{equation}

Where $E_{i_s}$ and $T_{i_s}$ are the shifted event-segment and the template-segment. $E_i$ is the $i^{th}$ event-segment composed of $E_i[0] = (t_i[0],v_i[0])$, and $E_i[1] = (t_i[1],v_i[1])$, $T_i[j]$ is the $j^{th}$ element of the $i^{th}$ template-segment, described by the tuple $(t,v) $. 

The deformation is then performed by computing a warped template segment as in Eq.~\ref{eq:segment_warp}:

\begin{equation}
\begin{split}
\label{eq:segment_warp}
   Time[j] &=T_{t,i_s}[j]\frac{E_{t,i_s}[1]}{T_{t,i_s}[L]} \\
   Value[j] &= T_{v,i_s}[j]+Time[j]\frac{E_{v,i_s}[1]-T_{v,i_s}[L]}{Time[L]}
\end{split}
\end{equation}

\begin{figure}[!t]
	\centering
    \includegraphics[width=0.8\linewidth]{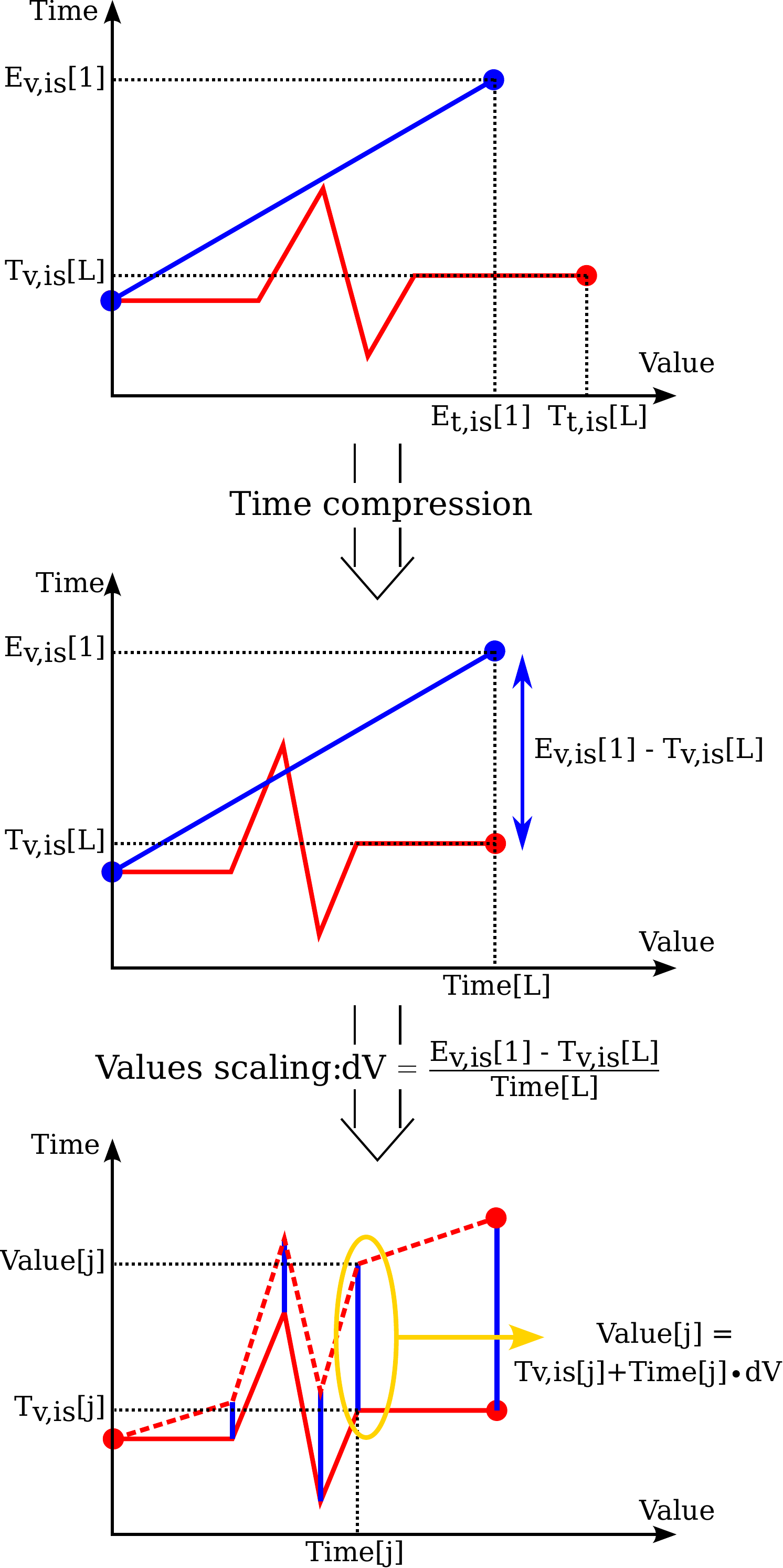}
	\caption{Exemplification of warping based on II-DDTW. In blue, a segment defined by two events, in red, the corresponding template segment.}
	\label{fig:segment_warping}
\end{figure}

This operation is graphically depicted in Fig.~\ref{fig:segment_warping}, aligning the edges of the template and event segments, while smoothly varying the values of the template segments by the same $\Delta v = \frac{E_{v,i_s}[1]-T_{v,i_s}[L]}{Time[L]}$ for each time-unit. 
The operations showed in Eq.~\ref{eq:segment_warp} compute a sequence of $(t,v)$ vectors uniformly sampled but not with the same time-base as the original signal and segment-dependent. Finally, the heartbeat is reconstructed by concatenating all the warped segments (adding the offset subtracted in Eq.~\ref{eq:shift_segment}) and linearly resampling the data to obtain a uniform time-base.

\subsection{Templates computation}
\label{subsec:templates_computation}

The templates computation is divided into five stages: re-computation trigger, U.S. Beats acquisition, clustering, templates filtering, and set re-computation.

\textbf{Templates set re-computation trigger.} 
After a new templates set is computed, the template computation block (Fig.~\ref{fig:system}-D) collects a reference set of 400 DTW distances from the results of Section~\ref{subsec:DTW}, assuming that, in this time span, the templates set is representative of the patient heartbeat morphologies. 
Hence, the distance set acquired is used as a reference to estimate the distribution of the DTW distance of a coherent templates set.
After the initial training time, the template computation block acquires a test batch of distances every 60 seconds and compare it with the reference distribution using the A.D. test~\cite{AD_test}. 
This test returns a critical value bounded by the probability that the test sequence came from the same distribution of the reference sequence. If such probability is lower than 0.05 for two consecutive times, then a new templates set is computed. This double confirmation is required to avoid triggering the templates set re-computation if the signal in one time window was affected by external factors that do not modify the beats morphology but might modify the recording (e.g., motion noise).

\textbf{U.S. Beats acquisition.} When a new templates set is needed, the first step is to start the signal acquisition at a uniform sampling rate. 
The uniformly sampled signal is then divided by the beats-division algorithm in U.S. beats. 
The first time the templates set is computed, the U.S. beats acquisition is performed for a longer time window (compared to templates set re-computation during normal functioning). 
Each heartbeat is then min-max normalized and used as input for the clustering algorithm (Fig.~\ref{fig:system}).

On an opposite note, during the re-computation stage, the acquisition time is shorter. This short acquisition time is required since this operation is triggered when the morphology of the current heartbeats do not resemble the ones contained in the already present templates set. The processing pipeline, hence, needs to quickly adapt to signal changes and a search-space small enough to consider the new morphologies as separated clusters and not outliers of already present and dominant ones.



\textbf{Self-organizing clustering.} The templates set is computed by clustering the acquired heartbeats and selecting the U.S. beats closest to the centroids. 
The choice of the clustering method is then driven by three main facts: 1) the number of clusters can not be determined beforehand. While it is true that a heartbeat can assume only a finite number of morphologies [citation needed], this would require the U.S. beats set under analysis to be statistically complete and without biases, and this can not be ensured for any subset of collected beats. 2) We can not ensure the clusters are convex for any given metric. Also, since our algorithm iteratively computes local clusters, their distribution could change between iterations, making convex conversions techniques such as the kernel trick~\cite{kernel_trik} not viable. 3) The clusters centroids need to be elements of the clustered set, since we can not ensure a heartbeat defined as an aggregation of examples to be physically coherent. 

These constraints are satisfied by graph-based clustering methods~\cite{Graph_clustering}.
Hence, our framework uses a self-organizing, graph-based clustering algorithm called affinity propagation~\cite{Affinity_propagation}. This algorithm requires two sets of parameters: graph distances and preferences.
The preferences vector is equivalent to a statistical prior that defines the likelihood of a data point to be a centroid. 

Our system uses the DTW measurement as the graph distance to capture the concept of morphological similarity between the dataset points. We then use the same preference value for each dataset point since we want the representative heartbeats to be emerging from the graph itself, without any external bias.

\textbf{Centroids filtering.} The centroids and relative clusters are filtered to satisfy two criteria: cluster dominance and the smallest possible signal-to-noise ratio (SNR). 
The first filter removes clusters that do not contain enough heartbeats. This threshold is 5\% of the total number of points in the dataset. 
In this scenario, the outliers to the clustering algorithms are the noisy portion of the signal. 
The second filter searches in each of the not discarded clusters a data point that is both the nearest possible to the centroid and with an SNR bigger than 17dB (50 in the linear scale).
The values used for these two filters were empirically computed with the objective of being very selective, and yet always allowing a non-zero amount of cluster and centroids in the small signal section analyzed.

To calculate the SNR value, the templates set computation stage first estimate the true signal and then the over-imposed noise. 
Then, assuming the noise to be a zero-median signal, it divide the signal from the noise applying a median filter to the acquired signal. 
This approach, however, requires the median filter to be as long as possible and short enough to not classify main signal features as noise. 
Given these conditions, the median filter time-span is set to 24 ms, based on the average duration of the QRS complex in an average heartbeat~\cite{SORNMO2005411}. 

\textbf{Templates set update.} To create an updated version of the templates set, we compare the old and the newly computed templates. Algorithm~\ref{alg:templ_update} details the templates set update process. The $update\_templates\_set$ procedure inputs are the old templates set ($old\_t$, list of templates) and the newly computed clusters ($new\_t$, a data structure holding the new templates, clusters centroids, and distances between all cluster elements and relative centroid). Moreover, the $dist\_from\_centroid$ (lines 7 and 16), computes the DTW distance between a template and a centroid. 

In Algorithm~\ref{alg:templ_update}, lines from 4 to 11 check if old templates are represented in the new clusters: first, line 5 fetches the nearest cluster to the old template under analysis, then, line 6 defines the threshold below which the old template is considered part of the new cluster. This threshold is the average DTW distance plus one standard deviation between the centroid and the elements of the cluster. If the old template is considered inside its nearest cluster, it is added to the list of possible representatives (line 9). Otherwise, it is kept (line 11) since it represents a morphology not captured by the current clustering iteration.
Lines 12 to 22 check if a new cluster is either a) represented by an old template (line 18), b) represented by the newly identified one (line 20), or c) if it represents a new morphology (line 22).

\begin{algorithm}[t]
\small
\caption{Templates set update algorithm}\label{alg:templ_update}
\begin{algorithmic}[1]
\Procedure{$update\_templates\_set$}{$old\_t,new\_t$} 
\State $near\_templates \gets Dict(List)$
\State $new\_set \gets List$

\For{$old \in old\_t$}
    \State $near\_new = get\_nearest(old,new\_t)$
    \State $threshold = cumpute\_th(near\_new.all\_dists())$
    \State $d = dist\_from\_centroid(old,near\_new)$
    \If{$d \leq threshold$}
        \State $near\_templates[near\_new.id] \gets old$ 
    \Else
        \State $new\_set \gets old$  \Comment{\textbf{Keep old}}
    \EndIf
\EndFor

\For{$new \in new\_t$}
    \If{$new.id \in near\_templates.id$}
        \State $old\_candidates = near\_templates[new.id]$
        \State $near\_old = get\_nearest(new,old\_candidates)$
        \State $d = dist\_from\_centroid(near\_old,n)$
        \If{$d \leq new.dist$}
            \State $new\_set \gets near\_old$  \Comment{\textbf{Keep old}}
        \Else
            \State $new\_set \gets new$  \Comment{\textbf{Update old}}
        \EndIf
    \Else
        \State $new\_set \gets new$  \Comment{\textbf{Insert new}}
    \EndIf
\EndFor

\Return $new\_set$
\EndProcedure
\end{algorithmic}
\end{algorithm}




\subsection{Processing pipeline composition}
\label{subsec:pipeline_composition}
The run-time behaviour of the framework depicted in Fig.~\ref{fig:system} and detailed in the previous sections can be summarized as follows:
\begin{enumerate}
    \item As the pipeline execution starts, no templates set is defined. The re-computation trigger initializes the templates acquisition block. This computation lasts for a defined but parametric amount of time.
    \item The beats division block divides the uniformly sampled signal into heartbeats. Each heartbeat is sent to the clustering algorithm block.
    \item The centroids computed by the clustering algorithm get filtered as described in Section~\ref{subsec:templates_computation}. As no previous templates set exist, the obtained centroids are not compared to any previous template.
    \item When the templates set computation is ready, the system starts the event-based signal acquisition.
    \item The EB-gQRS~\cite{EB_QRS} algorithm finds the QRS complexes in the event-based signal and, consequently, subdivide the EB signal in EB heartbeats.
    \item The II-DDTW algorithm in the DTW matching block matches any EB heartbeat to every template in the temples set. It then selects the template with the minimum distance from the analyzed EB heartbeat and marks it as representative of this heartbeat.
    \item For each EB heartbeat, the DTW matching block matches and warps template segments accordingly to the recorded events and the warping path obtained from the II-DDTW algorithm.
    \item Each warped segment is then recomposed together by the template warp block to obtain the reconstructed heartbeat.
    \item In order to find if a new templates set is needed, the A.D. Test block saves back the warping distances between the EB heartbeats and the selected representative template.
    \item After a new templates set is computed, the A.D. Test block acquires a vector of DTW distances as reference, assuming that a newly computed template is representative of the heartbeats generated immediately after the set computation.
    \item After defining the reference distance vector, the A.D. Test block acquires a new test vector and checks if it comes from the same probability distribution.
    \item In case the A.D. Test fails ($p-value \leq 0.05$) for two consecutive times the re-computation trigger signal starts a new templates set acquisition.
    \item Once a new templates set is computed, the clustering block compares it with the previous set and merges the two as seen in Section~\ref{subsec:templates_computation}.
\end{enumerate}

An implementation of this processing pipeline has been published as open-source\footnote{\url{https://c4science.ch/source/EB_ECG_Smart_Resampler/}}, using  Python as the primary language, with C implementations for the most compute-intensive routines such as II-DTW. The implementation is highly modular and parametric, allowing to freely vary the acquisition time for templates set computation, the lengths of the reference and test distance vectors, the clustering and centroids filtering parameters, the beats-subdivision window timing, and the number of points relevant for each template-segment/event bounding. The parameterization capability makes the algorithm suited for patient-specific tuning and application on different, non-ECG-related, self-similar signals that can be explored in future studies.



\section{Experimental setup}
\label{sec:setup}





\subsection{Data and methods}
\label{subsec:data}

To validate our work, we test it against the MIT-BIH Normal Sinus Rhythm database \cite{physioToolkit}. This database comprises 18 long-term (approximately one day) ECG recordings of subjects referred to the Arrhythmia Laboratory at Boston's Beth Israel Hospital (now the Beth Israel Deaconess Medical Center). The total number of heartbeats analyzed is approximately 1.8 million. The ECG recordings are sampled at a frequency of 128 Hz.



To decouple the performances of QRS detection from our algorithm, we relied on the ground-truth annotations in the selected dataset. This decision allows us to focus on the performance evaluation of the work developed here (the processing pipeline), effectively uncoupling our results from the performance of any QRS detection algorithm.

Using the QRS time locations, the processing pipeline computes the instantaneous RR interval and uses it to define the heartbeats window. The window boundaries for each heartbeat is defined by the time of the QRS complex minus 40\% and plus 60\% of the RR interval. Then, the acquisition block checks if any event is present at the boundaries time and, if not, it insert a synthetic event with a zero value. This approach is consequence of the boundary timing, as it is computed so to fall after a T-wave and prior to a P-wave, where no cardiac activity is present.

Finally, the initial long-term U.S. beats acquisition for the first templates set computation is set to 3 minutes, while the shorter time span for template-set re-computation is set to 40 seconds. These values have been empirically selected: a shorter acquisition time leads to less representative templates (undermining the results), and a longer time span do not lead to significant improvements.



\subsection{Evaluation metrics}
\label{subsec:res_eval_method}

The performances of the hereby proposed processing pipeline are evaluated using three merit figures: 1) percentage root mean square difference (PRD) \cite{RRMS}, 2) Dynamic time warping distance, and 3) P and T wave delineation $F_1$ score.
These three metrics complement each other,  depicting a comprehensive view of the strengths and weaknesses of our approach, as discussed in the following.

The PRD is a normalized distance between two vectors, computed as in Eq.~\ref{eq:rrms}, where $x_{org}$ and $x_{rec}$ are the samples of the original and reconstructed signal:

\begin{equation}
    \label{eq:rrms}
    PRD = \sqrt{\frac{\sum_{i=1}^{n}[x_{org}(i)-x_{rec}(i)]^2}{\sum_{i=1}^{n}[x_{org}(i)]^2}}
\end{equation}

The next metric explored broadens the definition of distance to a non-strictly mathematical one~\cite{distance_math} using the original DTW warping distance as a merit figure (i.e., $D_{N,M}$ in Eq.~\ref{eq:DTW}). This is the cumulative $\ell_1$ distance between elements in the optimal warping path between a reconstructed and the corresponding true heartbeat.

The final metric we analyze is wave delineation. A normal ECG heartbeat is composed of three main waves: P, T, and QRS. The QRS complex detection is, in this instance, a trivial task: it is a fundamental part of the processing pipeline and the clearest feature of an ECG signal. This feature is, hence, always present and correctly positioned. Conversely, the presence or absence of the P and T waves, alongside with their correct positioning, is a strong marker of well-reconstructed heartbeats.
To evaluate the wave presence/correctness, we first use the delineation tool \texttt{ECGPUWAVE} included in the \texttt{wfdb} software package~\cite{physioToolkit}. This tool identifies the P and T waves in a record. Then, we use the \texttt{bxb} tool, also contained in the \texttt{wfdb} software package, to find if the waves found in the original signal coincide with the waves found in the reconstructed heartbeats. Using the standards given by the \texttt{bxb} software, two waves are considered matching if their labels are, at most, 0.15 seconds apart.
This measure is numerically evaluated measuring the correctly detected waves ($True~Positive, TP$), the waves wrongly inserted by our approach ($False~Positive, FP$), and the waves our approach did not manage to recover ($False~Negative, FN$). Using this data, we compute the sensitivity score $S = \frac{TP}{TP+FN}$, and the positive predictive value $PPV = \frac{TP}{TP+FP}$. Sensitivity measures the percentage of waves in the original signal correctly reconstructed and well-positioned, while the positive predictive value $PPV$ measures the probability of a detected (reconstructed) wave to be present in the original signal and not be wrongly inserted. Finally, we compute the $F_1 = 2\frac{S\cdot PPV}{S+PPV}$ score: the harmonic mean between $S$ and $PPV$.
Since the labeling in the original dataset and in the reconstructed signal are computed using the \texttt{ECGPUWAVE}, the obtained $F_1$ score measures the difference when using a delineation algorithm on a uniformly sampled signal rather than on a reconstructed one.

While the PRD (which is a sum of element-wise distances) represents well the difference in values between two vectors, it also has intuition pitfalls. For example, both a constant signal and a slightly skewed but overall correct reconstructed heartbeat can result in the same PRD.

The DTW distance captures  the morphological distance of our results from the true signal, making this operation well suited for evaluating a reconstruction operation. Still, this approach can be misleading, mainly in discerning how a feature is better represented. For example, when approximating a quadratic function, a 2-pieces wise linear function might present a smaller DTW than a quadratic function well centered but with slightly higher values.

The PRD and DTW metrics give us an intuition on the correctness of the values of a reconstructed sequence, while the wave identification marks the presence of the feature of clinical interest. 
In summary, the correct positioning of such features is insufficient to deem the results correct, but a combination of correct wave positioning, standard DTW, and PRD values denotes a reconstruction where the main ECG elements are correctly placed, with similar values and morphology to the original waves.

\subsection{Baselines}
\label{subsec:baselines}
Alongside the developed pipeline results, we also compute the metrics mentioned above for the EB signal reconstructed through three standard resampling techniques. 
\begin{enumerate}
    \item Sample-and-hold method, where we hold the last recorded value in-between events. This technique represents the complete set of information obtained from the level-crossing technique sampling, as we have no means to know intermediate samples.
    \item Linear interpolation. This technique assumes a constant derivative between two consecutive points.
    \item Quadratic SP-line interpolation~\cite{b_spline_efficient}. This technique ensures a smooth interpolation.
\end{enumerate}

All the obtained results are evaluated against the EB sampling level, described by the number of bits used for the level-crossing algorithm (the number of levels being equal to $2^B$, where $B$ is the number of bits). However, this measurement does not directly represent the effect of the sampling on the signal. We tackle this problem by computing a compression figure called sampling reduction factor (SRF) for each EB sampling level. This metric is the ratio of discarded samples with respect to the total number of samples in the uniformly sampled signal.

To compute all the results mentioned above, we apply the developed processing pipeline to every heartbeat in every record in the dataset. We then compute each metric for each resulting reconstructed beat. Then, we re-evaluate the same metrics using the chosen resampling techniques. Finally, we derive the statistical distributions for both methods and metrics.



\section{Results and discussion}
\label{sec:results}

\begin{figure}[!t]
	\centering
    \includegraphics[width=\linewidth]{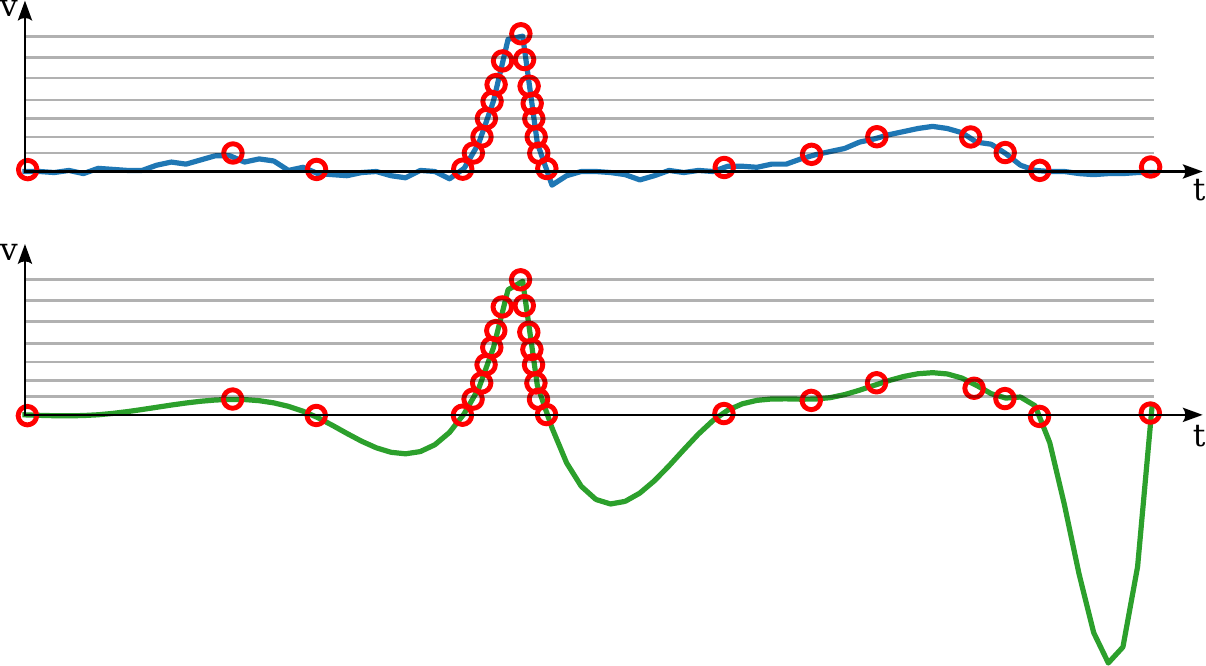}
    
	\caption{Spline interpolation of an event-based acquired heartbeat, using an LC-ADC with 6 bits of dynamic range. The non-uniform time distance between samples  forces the SP-line polynomial to be ill conditioned, exhibiting erratic behavior at the end of the signal or at the edges of the QRS complex.}
	\label{fig:sp_line}
\end{figure}

The analysis of the SP-line interpolation results shows, in the distance-based metrics, an error at least one order of magnitude higher than the other resampling techniques. As illustrated in Fig.~\ref{fig:sp_line}, the reason behind this behavior is the sparse nature of points in the initial and final section of every heartbeat together with high derivative values where the data-point density increases near the QRS complex. This behavior causes the polynomial approximation to be non-representative of the true underlying signal. This problem resides in the very nature of level-crossing sampling, making the SP-line interpolation a non-effective interpolation method for level-crossing ADCs. The results about SP-line interpolation are hence not discussed in the remaining part of this section while still being present in the numerical tabulated results.

\subsection{Sampling reduction factor}
\label{subsec:SRF_res}

\begin{figure}[!t]
	\centering
    \includegraphics[width=1\linewidth]{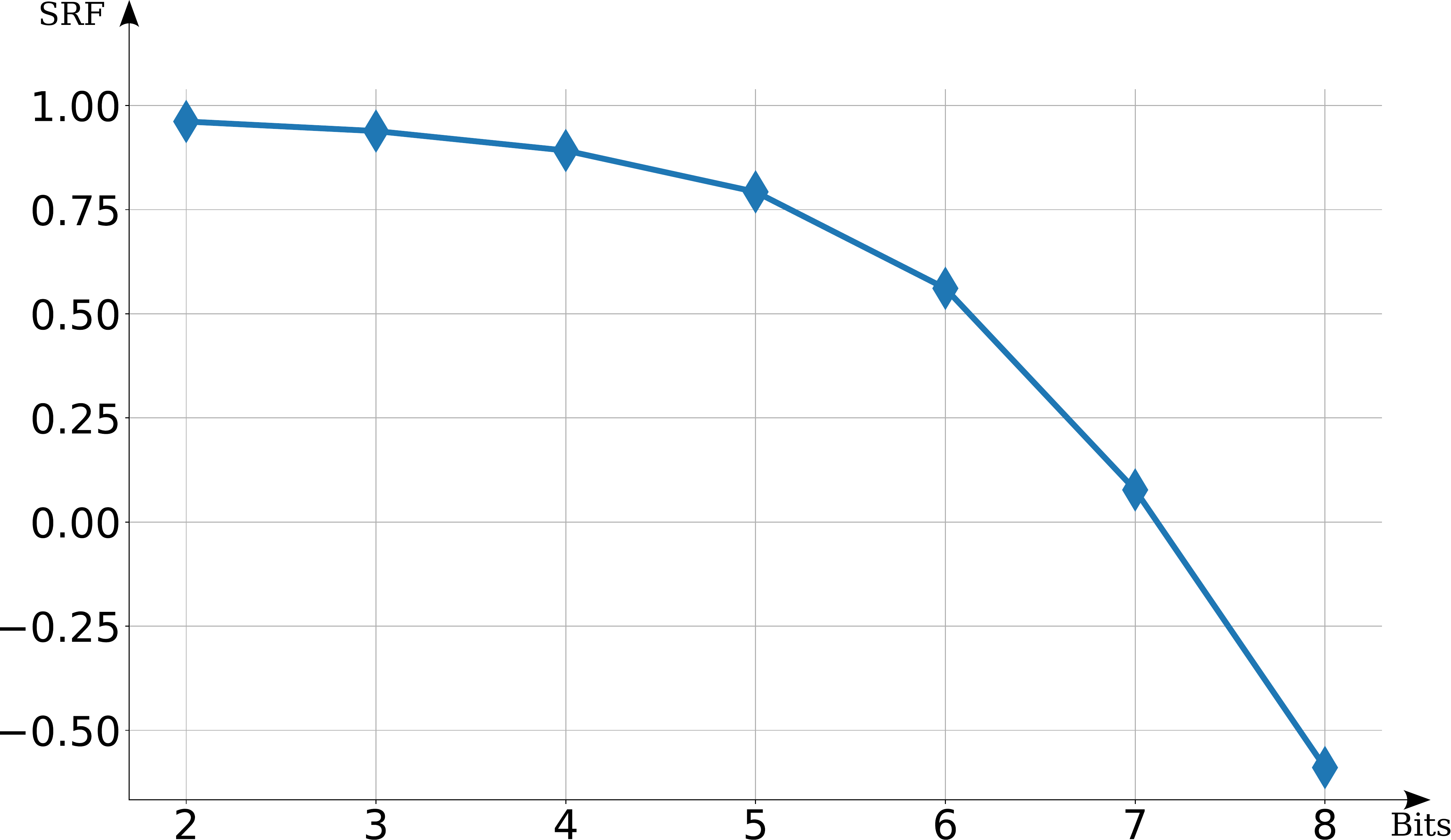}
    
	\caption{Sampling reduction factor vs. LC-ADC BIT numbers. Negative values represent a higher number of points with respect to uniform sampling.}
	\label{fig:LC_effects_SRF}
\end{figure}

Fig.~\ref{fig:LC_effects_SRF} shows the average SRF value, computed by EB-sampling all the signals in the dataset, for an increasing number of bits in the level-crossing EB-ADC. A negative SRF means that the number of points is higher than in the uniformly sampled signals.

The set of results presented in the next sections have been computed using 3, 4, and 5 bits. This decision is consequent to two observations: first, while using a 2-bits LC-ADC do not significantly lower the SRF, it fails to detect a single event in each heartbeat most of the times. This lead to results disconnected from any signal measurements. Second, the graph in Fig.~\ref{fig:LC_effects_SRF} shows, after 5 bits, a significant worsening of both the SRF value and derivative. Meanwhile, the results only show a marginal improvement for a 6-bits LC-ADC, while the results obtained for higher number of bits ($Bits \geq 7$) are indistinguishable from the uniform sampling approach.

Finally, we will focus on 4 bits results as we observe it to achieve a high SRF while obtaining overall good results.

\subsection{Percentage root mean square difference}
\label{subsec:PRD_res}

\begin{figure}[!t]
	\centering
    \includegraphics[width=0.9\linewidth]{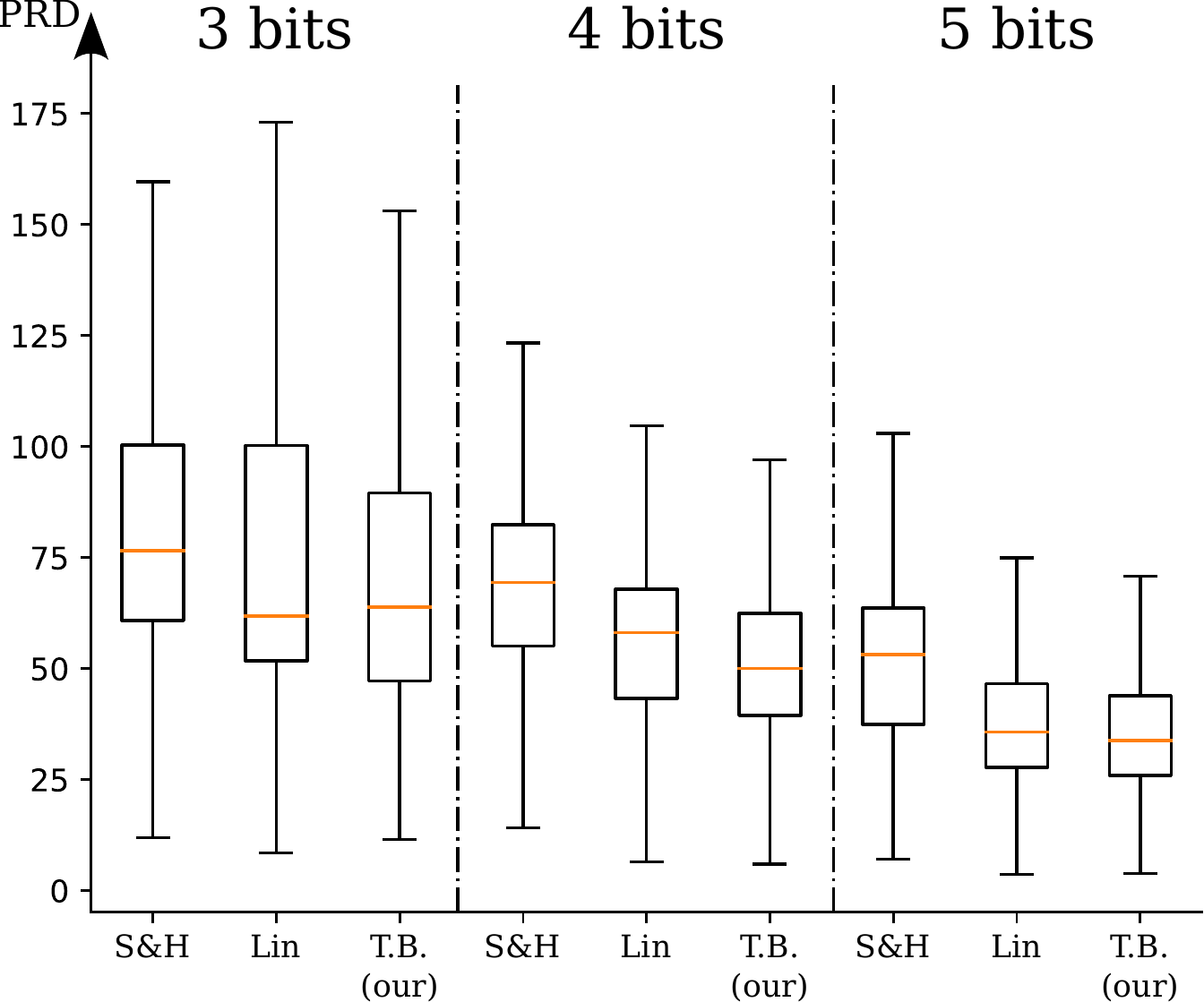}
    
	\caption{PRD distance distribution against LC-DC bit number and resampling/reconstruction method, using the number of bits discussed in Section~\ref{subsec:SRF_res}. S\&H: Sample\&hold resampling, Lin: Linear resampling, T.B. (our): template-based reconstruction (our method).}
	\label{fig:box_PRD}
\end{figure}

Fig.~\ref{fig:box_PRD} presents the distribution of the PRD distance of the three considered reconstruction techniques, using an LC-ADC with 3, 4, and 5 bits of dynamic range. The enclosed box spans from the 25th to the 75th percentiles, the horizontal central line is the median, and the thin outreaching lines enclose the 5-to-95 percentile.
Using this figure, we can derive two interesting points: 1) the flat interpolation is consistently outperformed by either the linear interpolation or the II-DTW method, 2) the II-DTW method is comparable to the linear interpolation, outperforming it when applied to a 4-bits LC-ADC.

The first point is the result of the step-defined function the sample\&hold method creates. This behavior causes, in fast varying signals, an increasing difference between the constant held value and the true underlying signal. The second point shows us that our technique is as valid as linear interpolation when we only consider samples magnitude.
The specific numerical values of the average PRD and standard deviation are shown in Table~\ref{tab:prd}.

\begin{figure}[!t]
	\centering
    \includegraphics[width=1\linewidth]{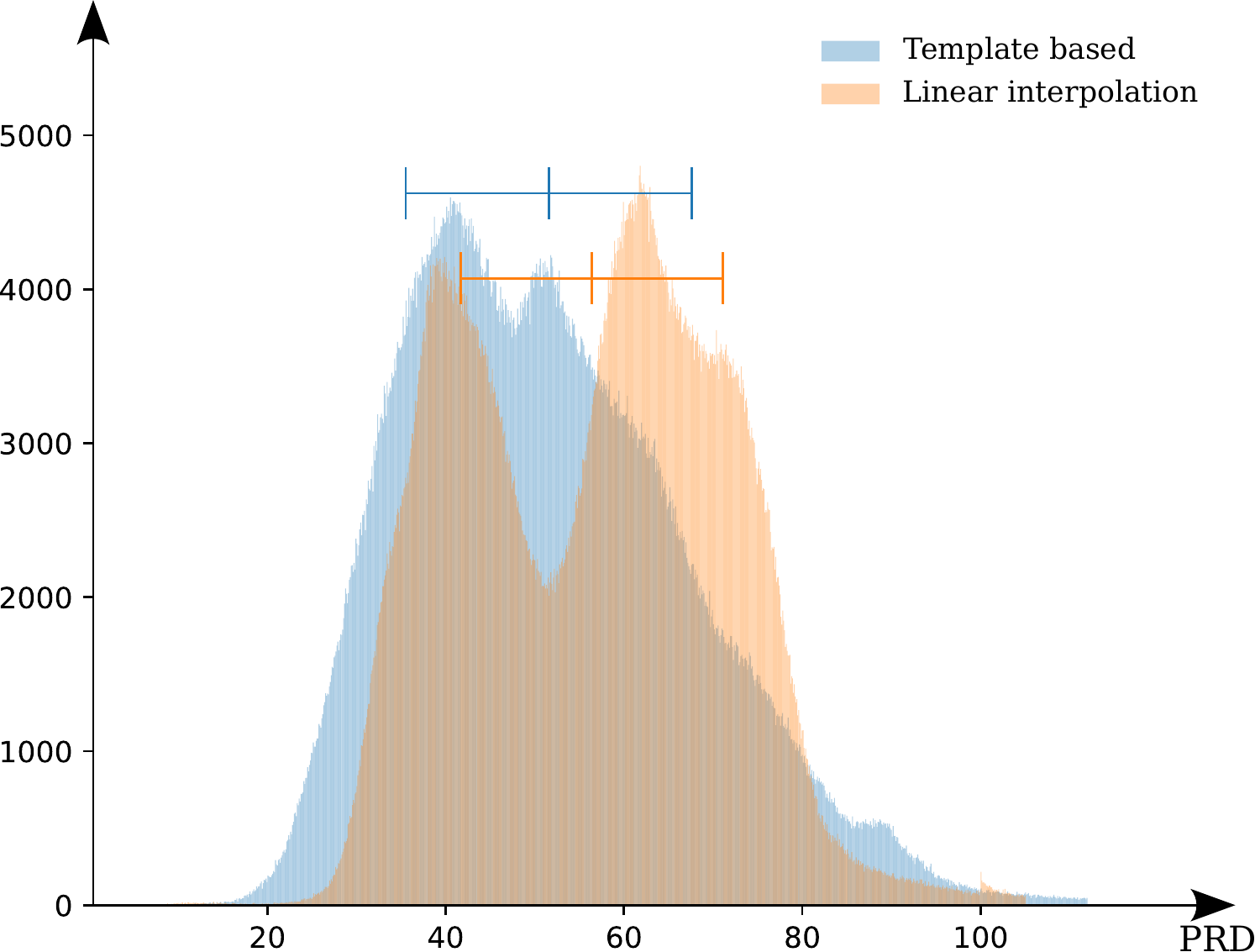}
    
	\caption{PRD distribution using an LC-ADC with 4 bits of dynamic range with highlighted average and standard deviation (horizontal bounded lines). Orange is Linear interpolation while blue is our proposed template based technique.}
	\label{fig:PRD_distribution}
\end{figure}

Fig.~\ref{fig:PRD_distribution} shows the PRD value distribution of the linear resample method and the II-DTW reconstruction using a 4-bits LC-ADC. The multi-modal distribution shown in the figure is caused by the different magnitudes of P and T waves. A small number of bits in the LC-ADC causes the distance between sampled levels to be sometime higher than the magnitude of the waves we desire to sample, with several possible behaviours: 1)~P-wave is sampled, T-wave not, 2)~P-wave is not sampled, T-wave is, 3)~both P and T-waves are sampled, 4)~both P and T-waves are not sampled. These conditions are less representative as the number of bits increases (and the span between ADC values diminishes).
Our methodology significantly mitigates this effect, explaining the better performances displayed in Fig.~\ref{fig:box_PRD} for LC-ADC with 4 bits of dynamic range.

\begin{table}[!t]
\caption{Average and standard deviation PRD for different resampling/reconstruction techniques and increasing EB-ADC bits.}
    \resizebox{\columnwidth}{!}{
    \begin{tabular}{l|r|r|r}
      \multicolumn{1}{c}{$ADC~bits~number$} & \multicolumn{1}{|c}{3} & \multicolumn{1}{|c}{4} & \multicolumn{1}{|c}{5} \\ \hline
      SP-line interpolation & $989\pm2993$ & $681\pm1505$ & $552\pm990$
      \\
      Sample \& Hold  & $79.2\pm25.0$ & $72.2\pm26.8$ & $52.3\pm16.9$  
      \\
      Linear interpolation & $71.3\pm27.1$ & $56.8\pm16.1$ & $37.9\pm14.3$
      \\
      \textbf{Template based (ours)} & $70.5\pm32.0$ & $52.7\pm19.8$ & $36.0\pm14.2$
    \end{tabular}
    }    
    \label{tab:prd}
\end{table}

\subsection{Dynamic time warping pseudo-edit distance}
\label{subsec:DTW_res}

\begin{figure}[!t]
	\centering
    \includegraphics[width=0.9\linewidth]{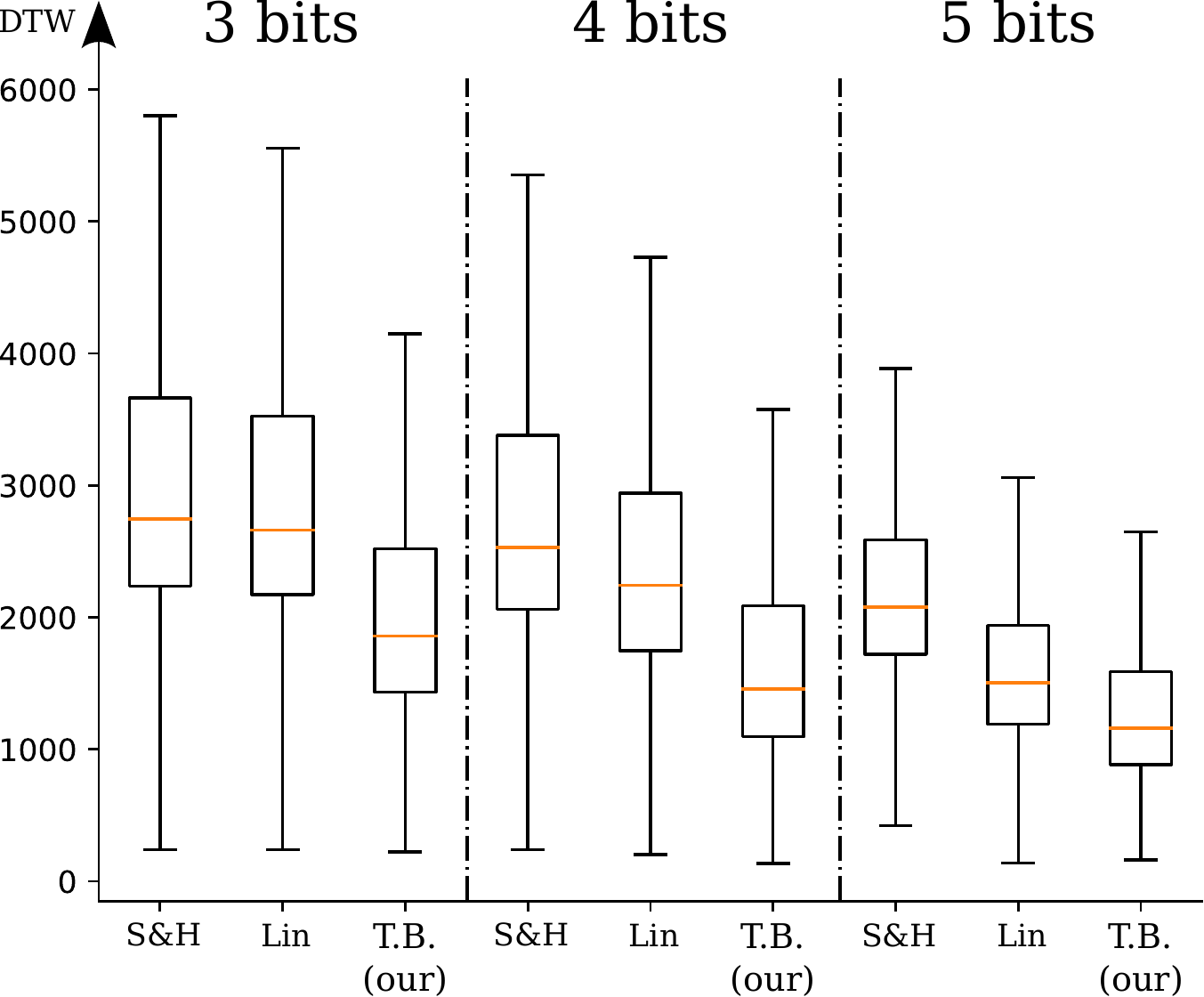}
    
	\caption{DTW distance distribution against LC-DC bit number and resampling/reconstruction method, using the number of bits discussed in Section~\ref{subsec:SRF_res}. S\&H: Sample\&hold resampling, Lin: Linear resampling, T.B. (our): template-based reconstruction (our method).}
	\label{fig:box_DTW}
\end{figure}

Fig.~\ref{fig:box_DTW} presents the distribution of the DTW distance of the three considered reconstruction techniques, using an LC-ADC with 3, 4, and 5 bits of dynamic range.
We observe that our proposed technique shows a consistently lower error for all choices of LC-ADC dynamic range.

\begin{figure}[!t]
	\centering
    \includegraphics[width=1\linewidth]{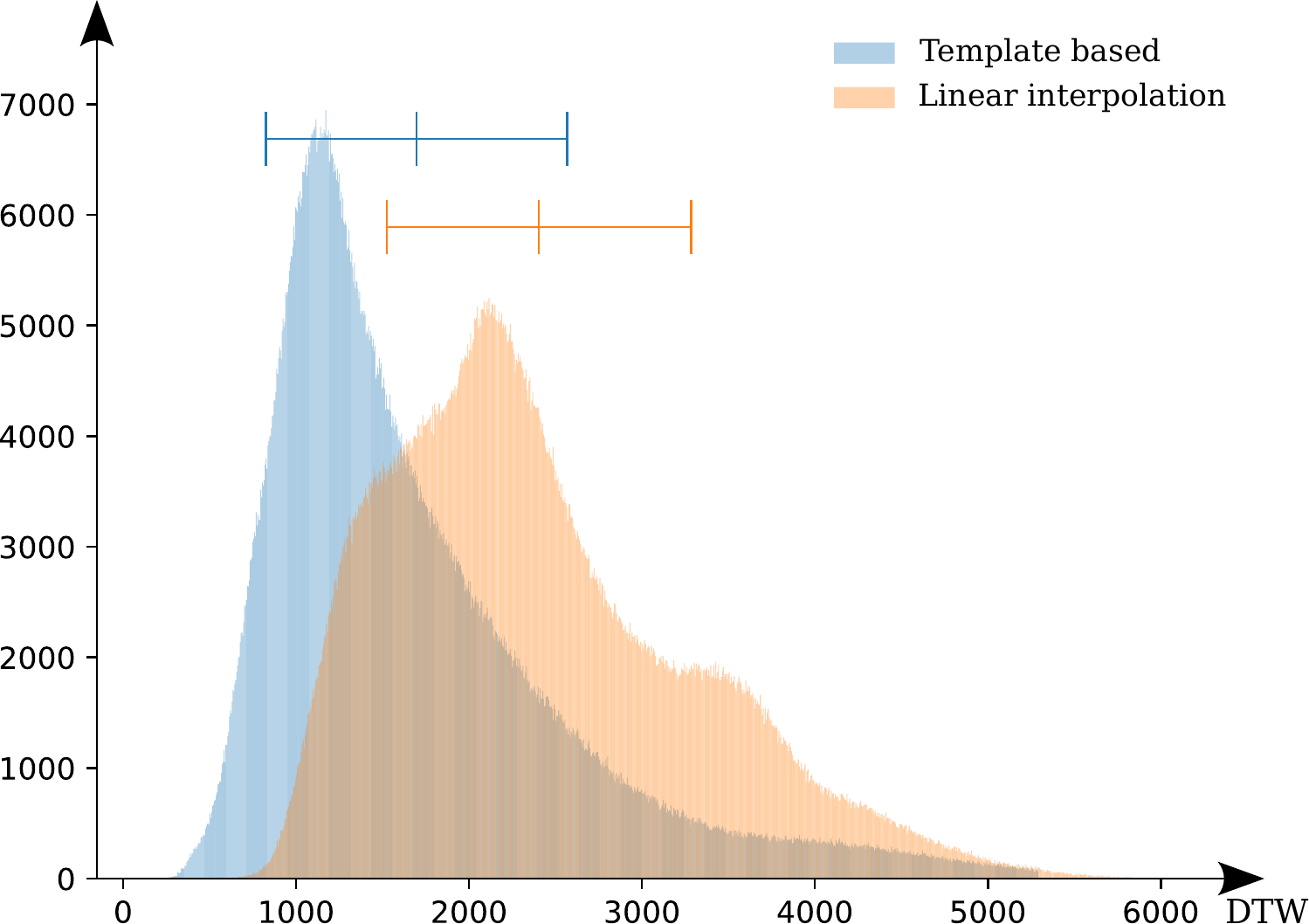}
    
	\caption{DTW distribution using an LC-ADC with 4 bits of dynamic range with highlighted average and standard deviation (horizontal bounded lines). Orange is Linear interpolation, while blue is our proposed template-based technique.}
	\label{fig:DTW_distribution}
\end{figure}

\begin{figure}[!t]
	\centering
    \includegraphics[width=0.9\linewidth]{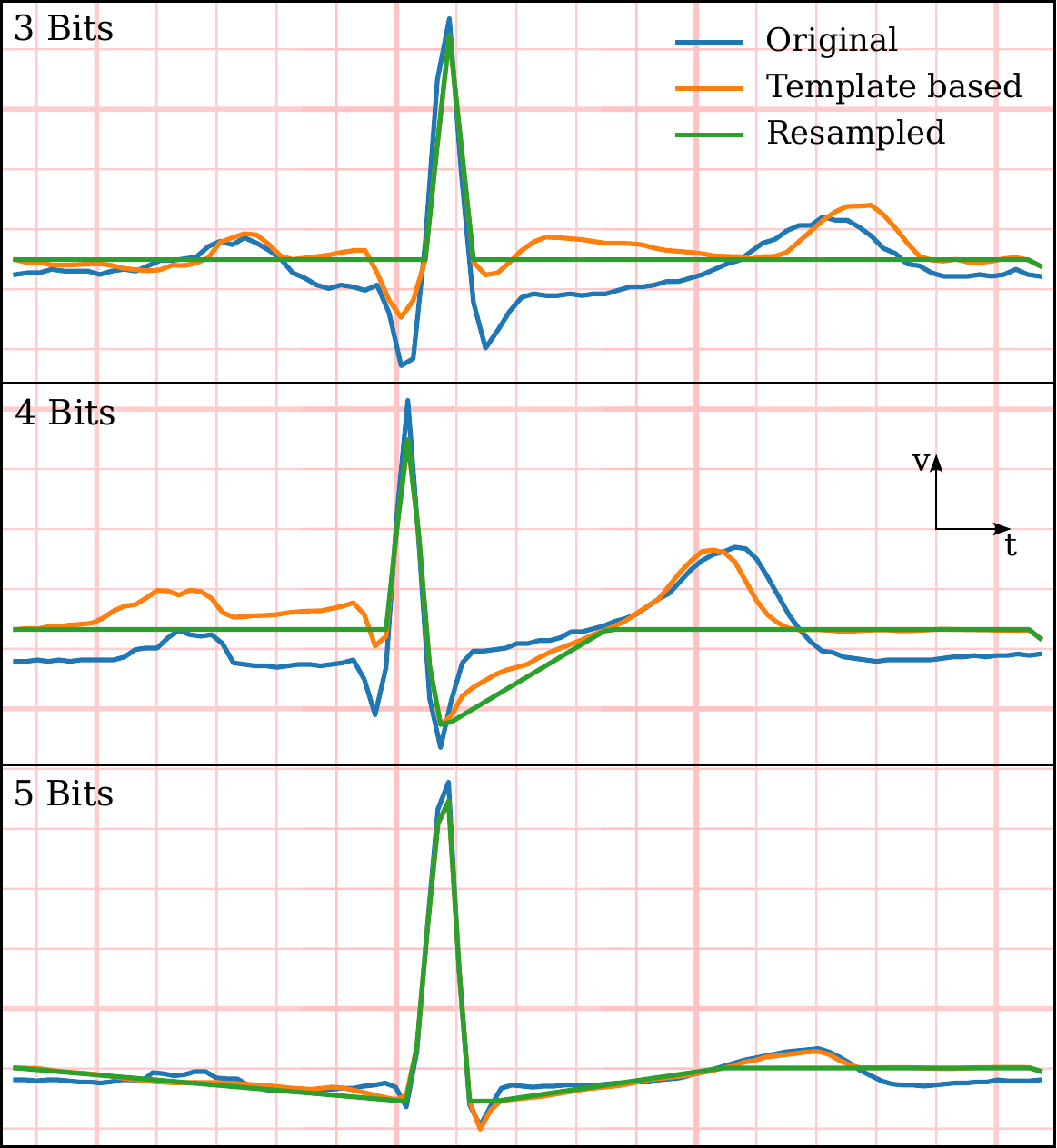}
    
	\caption{Heartbeat reconstruction associated with the $50^{th}$ DTW distance percentile, which includes an LC-ADC using 3, 4, and 5 bits of dynamic range.}
	\label{fig:DTW_perc}
\end{figure}

Fig.~\ref{fig:DTW_distribution} shows the detailed DTW distribution for the linear interpolation technique and our warping method for an LC-ADC with 4 bits of dynamic range. Here we can see differences in the distribution type between the two methods, with our warping reconstruction exhibiting a unimodal distribution while the linear interpolation method exhibits a composite behavior. These results can be explained by observing Fig.~\ref{fig:DTW_perc}, where we select, for all the analyzed levels, the heartbeat reconstruction whose DTW distance corresponds to the 50th percentile distance in the DTW distribution, behaving similar to what we would observe in an average scenario. The self-similarity prior knowledge allows our method to recover significant features also when no points are sampled in the section of interest, while the linear resampling technique  is not able to recover any information that has not been recorded.

Finally, Table \ref{tab:dtw} shows the average and standard deviation of the different reconstruction methods for the selected LC-ADC dynamic ranges. While showing an advantage of our method over classical resampling, these results highlight the accomplishment of the main objective of our work: a good morphological representation of an EB-sampled signal.

\begin{table}[!t]
\caption{Average and standard deviation DTW for different resampling/reconstruction techniques and increasing EB-ADC bits. The results are here scaled down by a factor $10^3$.}
    \resizebox{\columnwidth}{!}{
    \begin{tabular}{l|r|r|r}
      \multicolumn{1}{c}{$ADC~bits~number$} & \multicolumn{1}{|c}{3} & \multicolumn{1}{|c}{4} & \multicolumn{1}{|c}{5} \\ \hline
      SP-line interpolation & $38.4\pm36.6$ & $23.0\pm24.0$ & $15.2\pm19.2$
      \\
      Sample \& Hold & $3.05\pm2.07$ & $2.77\pm1.37$ & $2.22\pm0.90$
      \\
      Linear interpolation & $2.94\pm1.70$ & $2.42\pm1.14$ & $1.62\pm0.71$
      \\
      \textbf{Template based(ours)}& $2.16\pm1.69$ & $1.74\pm1.19$ & $1.32\pm0.78$
    \end{tabular}
    }    
    \label{tab:dtw}
\end{table}


\subsection{Delineation results}
\label{subsec:delineation_res}
Wave delineation is one of the main objectives in ECG signal analysis. This task consists in detecting the presence, position, and characteristics of P and T waves, alongside the QRS complex. Here, we focus on a sub-task of the wave delineation problem: p and T waves detection. 



Table \ref{tab:delineation} shows the delineation results for the proposed resampling methods and our technique, in terms of $F_1$ score. We can observe that our method outperforms any resampling technique for any selected LC-ADC range. This is due to the usage of templates that, if correctly chosen by the processing pipeline, contain similar P and T waves to the true signal, as shown in Fig.~\ref{fig:DTW_perc}. The presented waves, warped accordingly to the recorded events, closely resemble the original signal, making the delineation more effective than when applied to a resampled signal.

Finally, we noted that the increase in $F_1$ score is mainly caused by an increase in the sensitivity score $S$, and only marginally thanks to a bigger positive predictivity $PPV$ value. To exemplify this point, we observe P wave delineation, using a 4-bits LC-ADC, for linear resampling and templates-based reconstruction. The linear resampling achieve: $S = 0.019$, and $PPV = 0.630$, while using our method we measure: $S = 0.566$, and $PPV = 0.725$.

These results represent the core argument and contribution of our study. The DTW and PRD distances capture an overall improvement achieved by our method over classical resampling. However, they fail to reflect how this improvement is achieved or how it might be helpful for any task. The $F_1$ score in wave delineation shows that the improvements are caused by an effective reconstruction of the otherwise removed fundamental bio marker of the ECG signal. Moreover, we do not only show an effective reconstruction of the P and T waves but also their correct positioning with respect to the ground truth signal.

\begin{table}[!t]
\caption{P and T wave delineation results $F_1$ score for different resampling/reconstruction techniques and increasing EB-ADC bits}
\centering
    \resizebox{0.9\columnwidth}{!}{
    \begin{tabular}{l|r|r|r}
      \multicolumn{1}{c}{$ADC~bits~number$} & \multicolumn{1}{|c}{3} & \multicolumn{1}{|c}{4} & \multicolumn{1}{|c}{5} \\ \hline
      SP-line interpolation - P & 0.111 & 0.237 & 0.351
      \\ \cline{2-4}
      SP-line interpolation - T & 0.655 & 0.598 & 0.620
      \\ \hline
      Sample \& Hold - P & 0.009 & 0.051 & 0.101
      \\ \cline{2-4}
      Sample \& Hold - T & 0.231 & 0.394 & 0.697  
      \\ \hline
      Linear interpolation - P & 0.007 & 0.039 & 0.114
      \\ \cline{2-4}
      Linear interpolation - T & 0.231 & 0.353 & 0.617
      \\ \hline
      \textbf{Template based (ours) - P} & 0.646 & 0.696 & 0.699
      \\ \cline{2-4}
      \textbf{Template based (ours) - T} & 0.814 & 0.852 & 0.870
    \end{tabular}
    }    
    \label{tab:delineation}
\end{table}


\subsection{Different templates acquisition mode}
\label{subsec:templates_acq_mode}

As described in Section~\ref{sec:methods}, we compare the performance of our processing pipeline with two computationally lighter, open loop, variations of it: 1) the templates set is computed only at the beginning of the process, 2) only one template is acquired at the beginning of the process.

\begin{figure}[!t]
	\centering
    \includegraphics[width=0.9\linewidth]{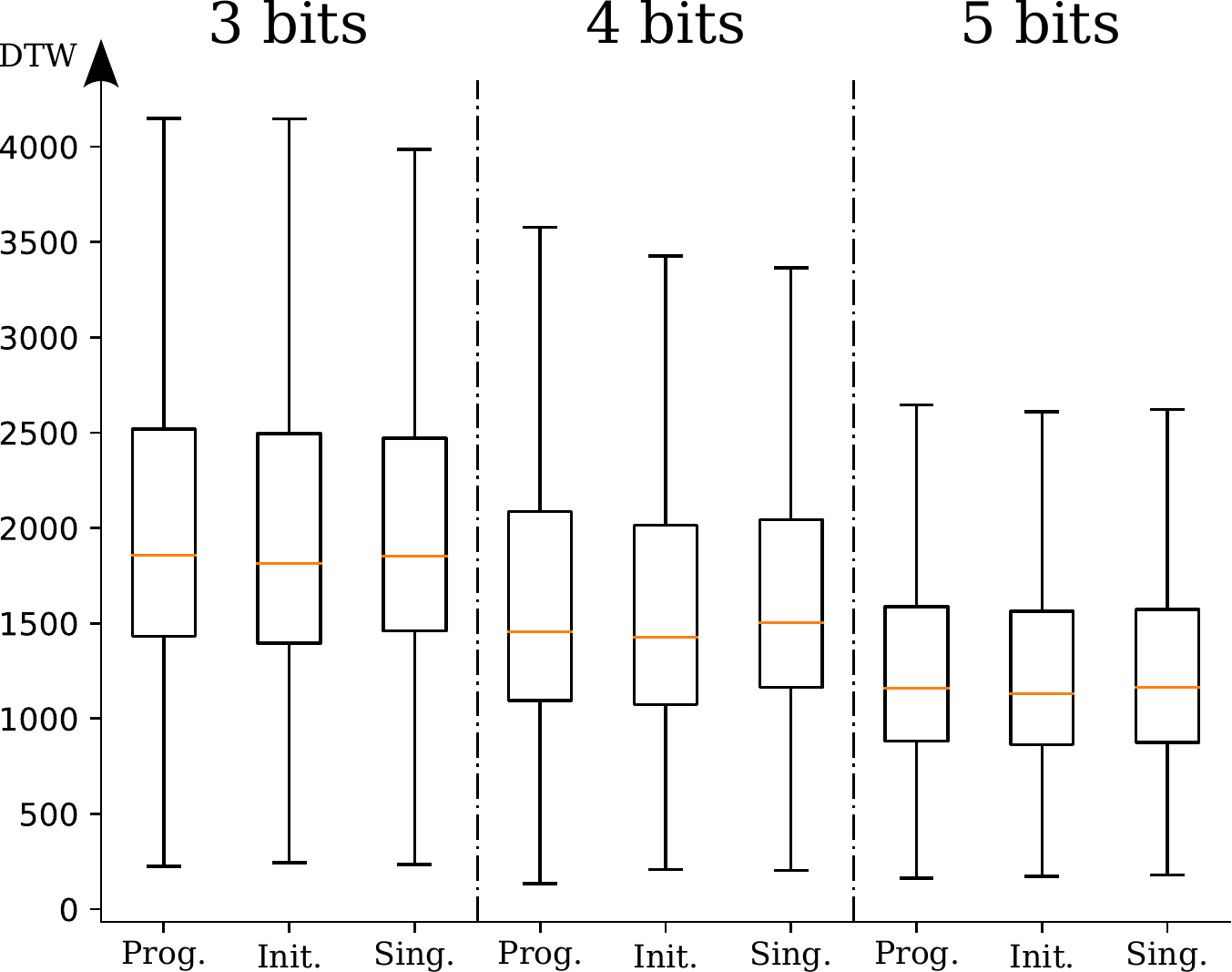}
    
	\caption{DTW distance distribution against LC-DC bit number and Templates acquisition type, using the number of bits discussed in Section~\ref{subsec:SRF_res}. Prog.: Progressive templates acquisition, Init.: Multiple initial templates, Sing.   : Single initial templates}
	\label{fig:box_DTW_templates}
\end{figure}
\begin{figure}[!t]
	\centering
    \includegraphics[width=0.95\linewidth]{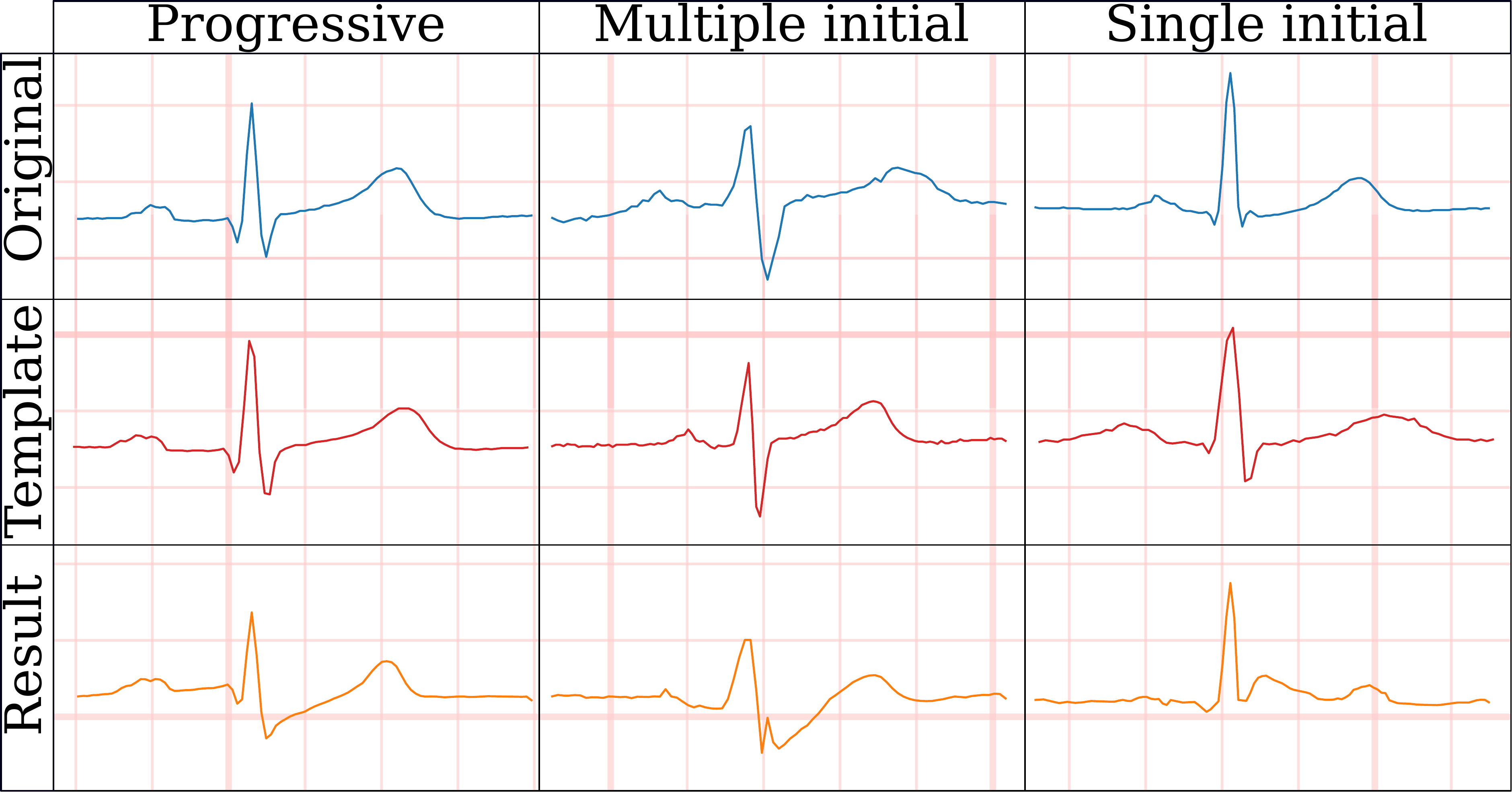}
    
	\caption{Heartbeat reconstruction associated with the $50^{th}$ DTW distance percentile, with an LC-ADC using 4 bits of dynamic range. Each column is a different templates-computation method. First column: progressive multi-template, second column: multiple templates acquired only at the beginning, third column: single initial template acquired at the beginning.}
	\label{fig:DTW_perc_templates}
\end{figure}

Fig.~\ref{fig:box_DTW_templates} shows the distributions of the DTW distance for the three proposed processing pipeline variations, using an LC-ADC with either 3, 4, or 5 bits of dynamic range. These results highlight a condition of indifference under this type of measurement. Moreover, an analysis of the previously explored merit figures (PRD and $F_1$ delineation score) also shows no statistically significant differences between the three reconstruction methodologies. However, as shown in Fig.~\ref{fig:DTW_perc_templates}, the observation of the heartbeats in the 50th percentile of the DTW distance distributions shows a progressive worsening in the reconstruction as we use less adaptive and lighter reconstruction techniques. 

Since every heartbeat of a patient shares common features~\cite{SORNMO2005411}, even only one generic heartbeat can be representative of a significant portion of a complete record. This makes the warping and the consequential main waves positioning as effective as using more specialized templates. 
However, as we show in Fig.~\ref{fig:DTW_perc_templates}, the lack of an adaptive technique to determine the best template for a set of events, and the absence of a mechanism to detect templates set significance, makes the solution not effective in representing the correct type of waves, especially in long-lasting recordings like the one obtained by a Holter ECG.

\section{Conclusions}
\label{sec:conclusions}
In this work, we have developed and implemented a signal processing pipeline able to reconstruct event-based~(EB) sampled electrocardiogram (ECG) signals, through patient-specific heartbeat templates. This is accomplished by computing a set of locally representative heartbeats, selecting the best fitting one for each event-based sampled heartbeat, and warping it accordingly to the recorded events.
We warp the templates using a novel formulation of the DDTW algorithm, named Information Injected-DDTW~(II-DDTW), which uses the timing information of each event to bias the distance metric.
The templates are dynamically re-computed whenever they stop being representative of the underlying physical signal, allowing the processing pipeline to select the most representative template for each patient and signal section.
When compared to standard resampling techniques~(i.e., SP-line, sample-and-hold, and linear interpolation), we have shown that our proposed pipeline obtains a 10x improvement in P-wave reconstruction, a 2x improvement in T-wave reconstruction, and a 30\% improvement, on average, in morphological similarity with the underlying physical signal.

\vspace{-2pt}
\section*{Acknowledgements}

This work has been supported in part by the Swiss NSF ML-Edge Project (GA no. 200020\_182009), the EC H2020 DIGIPREDICT Project (GA no. 101017915), and the  EC H2020 DeepHealth Project (GA no. 825111). T.T. is supported by a Maria Zambrano fellowship (MAZAM21/29) from the UPV/EHU and the Spanish Ministry of Universities, funded by the European Union-Next-GenerationEU.









\bibliographystyle{elsarticle-num}

\vspace{-10pt}
\bibliography{references}

\end{document}